\documentclass[aps,prb,showpacs,twocolumn,groupedaddress,superscriptaddress,amsmath,amssymb,nobibnotes]{revtex4-2}

\usepackage{blindtext}
\usepackage[colorlinks,
          linkcolor=blue,
            citecolor=blue,
            urlcolor=blue
           ]{hyperref}
\usepackage{graphicx}
\usepackage{color}
\usepackage{txfonts}
\usepackage{bm}
\usepackage{natbib}
\usepackage{float}
\usepackage{ulem}
\usepackage{chngcntr}

\begin{document}
\title{Emergent Intermediate Phase in the $J_1$-$J_2$ XY model from Tensor Network Approaches}

\author{Feng-Feng Song}\email{song@issp.u-tokyo.ac.jp}
\affiliation{Institute for Solid State Physics, The University of Tokyo, Kashiwa, Chiba 277-8581, Japan}

\author{Hanggai Nuomin}
\affiliation{Department of Chemistry, Duke University, Durham, North Carolina 27708, United States}

\author{Naoki Kawashima}\email{kawashima@issp.u-tokyo.ac.jp}
\affiliation{Institute for Solid State Physics, The University of Tokyo, Kashiwa, Chiba 277-8581, Japan}
\affiliation{Trans-scale Quantum Science Institute, The University of Tokyo, Bunkyo, Tokyo 113-0033, Japan}

\date{\today}

\begin{abstract}
We investigate the finite-temperature phase diagram of the classical $J_1$-$J_2$ XY model on a square lattice using a tensor network approach designed for frustrated spin systems. This model, characterized by competing nearest-neighbor and next-to-nearest-neighbor interactions, exhibits a complex interplay between $U(1)$ and $Z_2$ symmetries. Our study reveals an emergent intermediate phase around $J_2/J_1 \sim 0.505$, which is characterized by a $Z_2$ long-range stripe order without phase coherence in the XY spins. The intermediate phase features two well-separated phase transitions: a higher-temperature Ising transition and a lower-temperature Berezinskii-Kosterlitz-Thouless transition. The relative separation between these transitions is significantly larger than previously reported, enabling a clearer investigation of their distinct thermodynamic properties. For $0.5<J_2/J_1 < 0.501$, two transitions merge into a single first-order phase transition, a phenomenon that cannot be explained solely by mapping to the Ising-XY model. As $J_2/J_1 \to \infty$, the transition evolves continuously into the BKT universality class. These findings advance the understanding of the mechanisms driving phase transitions in frustrated spin systems and suggest potential experimental realizations in platforms such as ultracold atoms, Josephson junction arrays, and optical lattices.
\end{abstract}

\maketitle

\section{Introduction}
The concepts of phases and phase transitions lie at the heart of our understanding of the physical universe, serving as a cornerstone for modern condensed matter physics and statistical mechanics. Seminal examples include the two-dimensional (2D) classical Ising model, governed by discrete $Z_2$ symmetry~, and the 2D XY model, characterized by continuous $U(1)$ symmetry\cite{Berezinsky1970,Kosterlitz1973}. A particularly intriguing aspect of phase transitions arises from the interplay between symmetries and the frustrations caused by competing interactions~\cite{Lacroix2011,Diep2020}. Such interplay can give rise to a rich variety of phases and exotic transitions at finite temperatures. 

One of the most prominent examples exhibiting such behaviors is the 2D fully frustrated XY model (FFXY) with competing ferromagnetic and antiferromagnetic interactions~\cite{Villain1977}. Although the model is $U(1)$ invariant, a new $Z_2$ degree of freedom emerges as a result of the minimization of conflicting local interactions. Over the past few decades, significant theoretical efforts have been devoted to studying the 2D FFXY model. However, the exact nature of the phase transitions remained elusive, largely due to the extremely close proximity of two distinct types of ordering~\cite{Teitel1983,Choi1985,Yosefin1985,Thijssen1990,Santiago1992,Granato1993,LeeJR1994,LeeSy1994,Santiago1994,Olsson1995,Cataudella1996,Olsson1997,Boubcheur1998,Hasenbusch2005,Okumura2011,Nussinov2014,Lima2019}. Recent studies have provided compelling evidence of two phase transitions at different temperatures, separated by an intermediate chiral long-range-ordered (LRO) phase in which phase coherence between XY spins is absent~\cite{Okumura2011,Song2022}. Unfortunately, the intermediate phase is extremely narrow, with the separation between the upper Ising transition and the lower Berezinskii-Kosterlitz-Thouless (BKT) transition being less than $2\%$.  The narrow separation has also been observed in triangular antiferromagnetic XY models~\cite{Miyashita1984,Shih1984,DHLee1984,DHLee1986,Korshunov1986,Himbergen1986,Xu1996,LeeSy1998,Luca1998,Song2023}. Such narrow separation makes experimental realization and detection challenging~\cite{Struck2013,Cosmic2020}.

To better understand the nature of these phase transitions, it is crucial to achieve a larger separation between them to reduce proximity effects. It is possible to realize a large separation as shown in the study of the Ising-XY model,  which explicitly incorporates interactions between $Z_2$ and continuous order parameters~\cite{Granato1987,Granato1991,Lee1991,Knops1994,Li1994,Nightingale1995,Lee1997,Cristofano2006,Ma2024}. The Ising-XY model is believed to belong to the same universality class as the FFXY model corresponding to a branch of simultaneous loss of Ising and XY order, as predicted by renormalization group analysis. While physically distinct, the Ising-XY model suggests that the Ising and XY transitions can be well separated by tuning the interaction between different types of topological defects, such as Ising domain walls and XY vortices~\cite{Granato1991}. 

A promising candidate for achieving such tuning is the 2D $J_1$-$J_2$ XY model on a square lattice, which introduces a large parameter space by varying the ratio of nearest-neighbor (NN) interactions $J_1$ and next-to-nearest-neighbor (NNN) interactions $J_2$. In this model, frustration arises from the competition between ferromagnetic NN and antiferromagnetic NNN interactions. Unlike the FFXY model, however, the discrete $Z_2$ ground-state degeneracy in the $J_1$-$J_2$ XY model originates from entropic thermal fluctuations rather than energetic contributions~\cite{Henley1989}, making it a prominent example of the ``order-by-disorder" effect. From the Coulomb gas formulation, the $J_1$-$J_2$ XY model was proposed to belong to the class of Ising-XY models, with the transitions corresponding to a specific branch of the critical line in the Ising-XY phase diagram~\cite{Simon1997Ising,Simon1997}. Despite a long history of investigation~\cite{Henley1989,Fernandez1991,Spisak1994,Simon1997,Loison2000}, the phase diagram of the $J_1$-$J_2$ XY model has remained only partially resolved. Renormalization group analysis initially suggested the possibility of a single phase transition separating the low-temperature $Z_2$-ordered phase from the high-temperature disordered phase~\cite{Henley1989,Simon1997}. Early Monte Carlo simulations found evidence for a single phase transition at $J_2/J_1 = 1$, with simultaneous Ising and XY ordering~\cite{Henley1989,Fernandez1991}. Later, more extensive Monte Carlo studies revealed two closely spaced transitions at $J_2/J_1 = 0.7$~\cite{Loison2000}. However, these results remain limited by critical slowing down and the effects of large frustration at low temperatures~\cite{Swendsen1987,Wolff1989,Rakala2017}. As a result, the precise structure of the phase diagram, especially in the regime near $J_2/J_1 \sim 0.5$, remains unresolved.

In this work, we carry out a detailed study of the $J_1$-$J_2$ XY model by systematically varying the ratio $J_2/J_1$ using a recently developed tensor network (TN) method for frustrated spin systems~\cite{Vanhecke2021,Song2022,Song2023kagome,Song2023}. The TN approach is particularly well-suited for investigating the low-temperature regime of frustrated models, as the massive degeneracy induced by strong frustration can be efficiently encoded in the local tensors. Moreover, the singularities of the entanglement entropy in the 1D transfer operator enable precise identification of phase transitions based on a unified criterion~\cite{Haegeman2017}. Thus, the complete phase diagram of the $J_1$-$J_2$ XY model is obtained. We first confirm the findings of previous studies at $J_2/J_1=1.0$ and $0.7$. Then, as $J_2/J_1$ decreases, we find an enlarged region of the intermediate phase with a relative separation between upper Ising and lower BKT transitions around $20\%$ at $J_2/J_1\sim0.505$. The intermediate phase is characterized by a $Z_2$ long-range stripe order in the absence of phase coherence between XY spins. The larger separation between the transitions enables us to resolve the distinct thermodynamic properties of the BKT and Ising transitions without the proximity effects that have hindered previous studies of the FFXY model. Consequently, the nature of the two transitions becomes much clearer. Interestingly, as $J_2/J_1$ decreases further below $0.501$, the two phase transitions merge into a single first-order phase transition, a phenomenon that has not been reported before. This finding reveals a new aspect of the $J_1$-$J_2$ XY model and highlights the interplay between different degrees of freedom in determining the nature of phase transitions.

The rest of the paper is organized as follows. In Sec.~\ref{sec:model}, we introduce the $J_1$-$J_2$ XY model and discuss the order-by-disorder effects. In Sec.~\ref{sec:tensor}, we describe the application of the tensor network method to this frustrated system. In Sec.~\ref{sec:result}, we present the numerical results, construct the finite-temperature phase diagram, and analyze the properties of the different phases. Finally, in Sec.~\ref{sec:conclusion}, we summarize the main findings of this study and discuss their implications.

\section{$J_1$-$J_2$ XY Model}\label{sec:model}
\begin{figure}[tbp]
    \centering
    \includegraphics[width=0.99\linewidth]{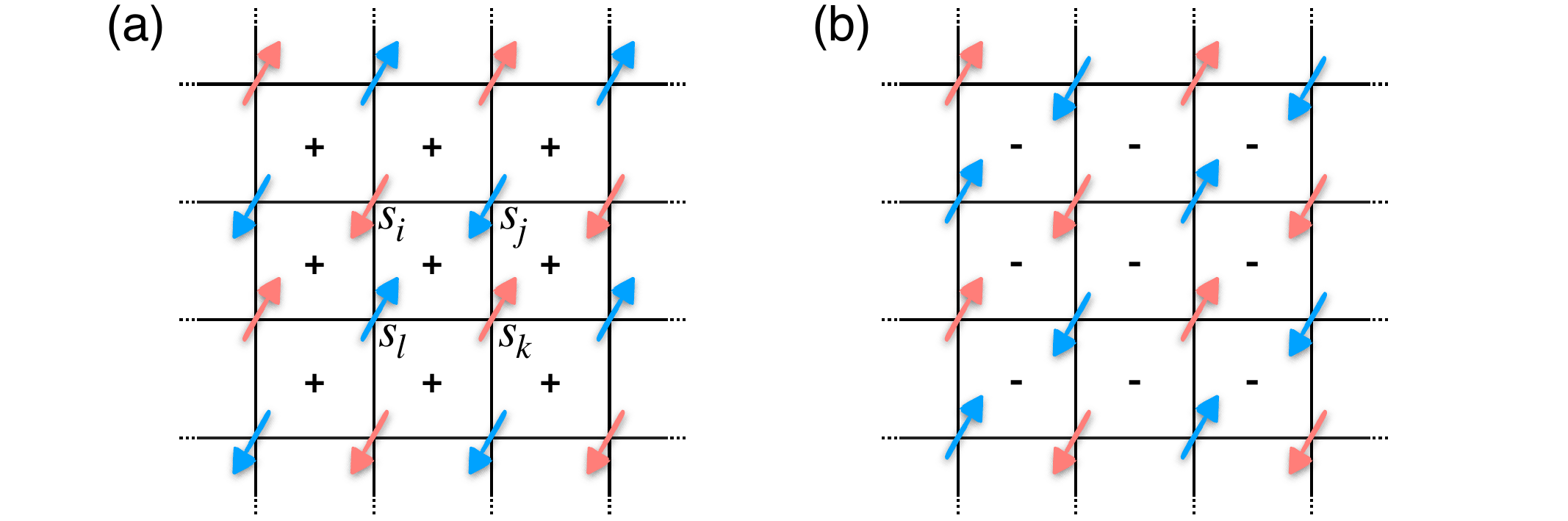}
    \caption{
    Two degenerate ground state configurations for the $J_1$-$J_2$ XY model with $J_2/J_1>0.5$, where red and blue colors denote antiferromagnetic sublattice $A$ and $B$.
    (a) Horizontal stripe order with chirality $\tau=+1$ on each plaquette. 
    (b) Vertical stripe order with chirality $\tau=-1$ on each plaquette. }
    \label{fig:model}
\end{figure}

The Hamiltonian of the $J_1$-$J_2$ XY model is defined on a 2D square lattice as
\begin{equation}
    H=-J_1\sum_{\langle i,j\rangle}\cos(\theta_i-\theta_j)+J_2\sum_{\langle\langle i,j\rangle\rangle}\cos(\theta_i-\theta_j),
\end{equation}
where $J_1$ and $J_2>0$, $\theta_i$ denotes the spin orientation at the lattice site $i$ and the sums of  $\langle i,j\rangle$  and $\langle\langle i,j\rangle\rangle$ run over all NN and NNN sites, respectively.

When $J_2/J_1<0.5$, the ground state is ferromagnetic with $U(1)$ degeneracy. As the temperature increases, there will be a BKT transition driven by the dissociation of vortex pairs. From the Coulomb gas formulation, the critical temperature should be approximately rescaled by a factor of $1-2J_2/J_1$ compared to the original XY model with only NN interactions~\cite{Simon1997Ising}.

When $J_2/J_1>0.5$, the ground state breaks into two square $\sqrt{2}\times\sqrt{2}$ sublattices with antiferromagnetic order. As shown in Fig.~\ref{fig:model}, the red and blue spins denote two antiferromagnetic sublattices, $A$ and $B$. The ground state energy density,  $e_0=-2J_2$, does not depend on the relative orientation $\phi=\theta_A-\theta_B$ between two sublattices, implying a $U(1)\times U(1)$ degeneracy. However, this degeneracy is lifted by thermal fluctuations through the well-known ``order by disorder" mechanism~\cite{Henley1989}. Thermal fluctuations favor collinear ordering with $\phi = 0$ or $\phi = \pi$, reducing the symmetry to $U(1) \times Z_2$. The details of this effect are discussed in Appendix~\ref{sec:spinwave}, where we show that small spin-wave fluctuations lower the free energy by an amount proportional to $0.32(J_1\cos\phi/2J_2)^2$. Two degenerate ground states are shown in Fig.~\ref{fig:model} with a stripe order in horizontal and vertical directions, respectively. The $Z_2$ stripe order can be characterized by the chirality defined on each plaquette
\begin{equation}
\tau_i=\frac{1}{4}(\bm{s}_i-\bm{s}_k)\cdot(\bm{s}_j-\bm{s}_l),
\label{eq:tau}
\end{equation}
where sites $i$, $j$, $k$, and $l$ are four corners of a plaquette with diagonal $(i,k)$ and $(j,l)$ as shown in Fig.~\ref{fig:model} (a). Then $\tau=\pm1$ corresponds to horizontal and vertical stripe orders, respectively.

From a renormalization group analysis, the $Z_2$ symmetry-breaking effect from spin waves is strong and remains relevant even in the limit of $J_2/J_1 \gg 1$~\cite{Jose1977,Simon1997}. In the strong coupling limit, the two sublattices become locked in an Ising-like manner, and the critical behavior of the $J_1$-$J_2$ XY model is described by a coupled double Coulomb gas, which falls within the universality class of Ising-XY models. From this mapping, the transition in the $J_1$-$J_2$ XY model was proposed to correspond to a part of the critical line between the branch and tricritical points in the Ising-XY model. The interplay of the $U(1)$ and $Z_2$ order parameters can result in unusual critical behavior, where both Ising and XY orders vanish simultaneously. However, this scenario remains controversial, as it is unclear whether the system undergoes two closely spaced transitions or a single unified transition. Due to the limited simulations available for this model, definitive conclusions are difficult to achieve~\cite{Fernandez1991,Loison2000}. As far as we know, the precise nature of the phase transitions remains partially explored.

\section{Tensor network methods}\label{sec:tensor}
\begin{figure}[tbp]
    \centering
    \includegraphics[width=0.99\linewidth]{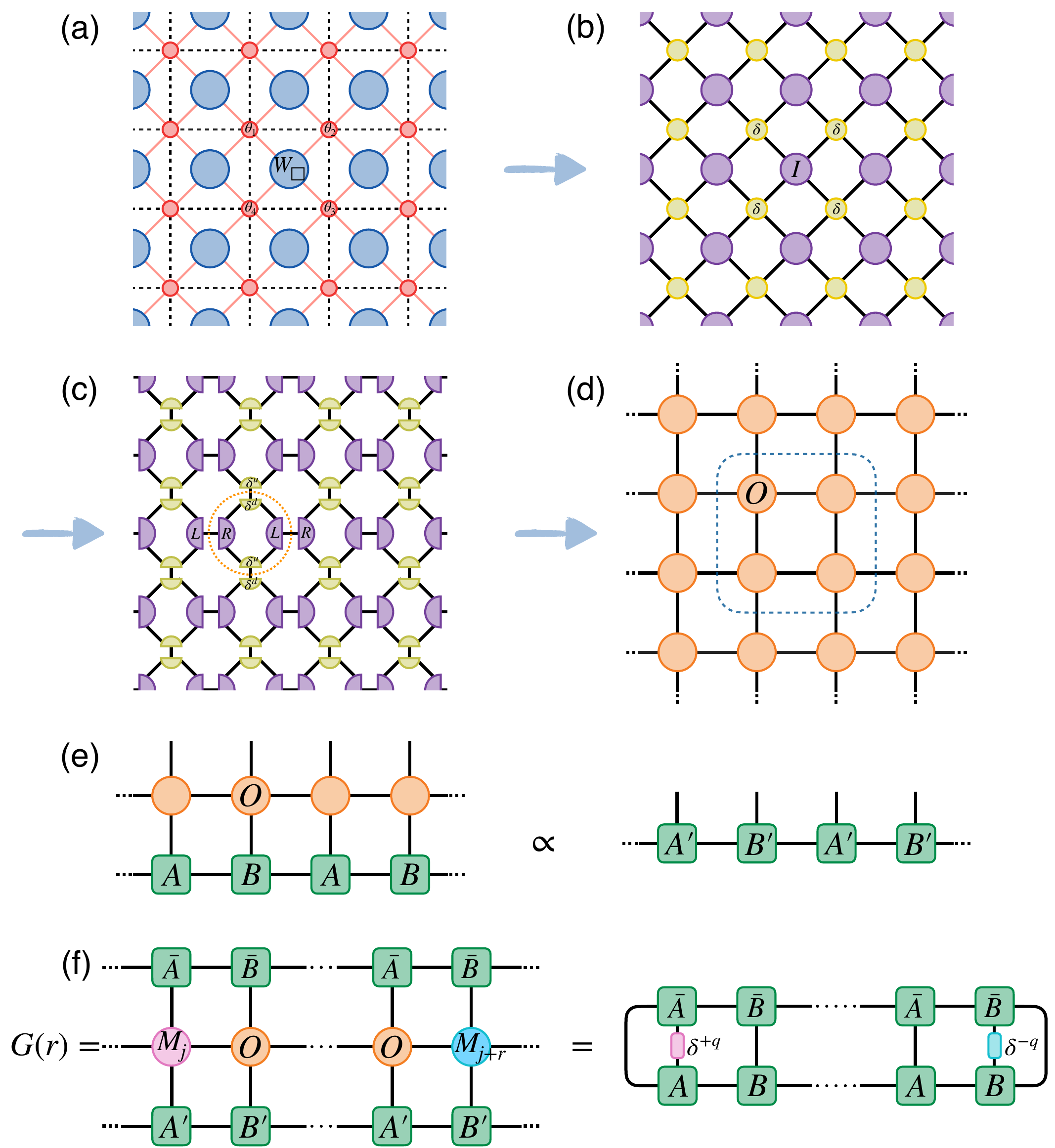}
    \caption{
    Tensor network representation of the $J_1$-$J_2$ XY model on a square lattice. 
    (a) The tensor network with continuous indices, where the $W$ tensors account for the Boltzmann weight on each plaquette, and the red dot tensor represents the integration of the joint $\theta$ variables among four squares. The black dotted line denotes the original square lattice.
    (b) The tensor network with discrete indices obtained from Fourier transformations. The $I$ tensors carry the Boltzmann weight and the $\delta$ tensors ensure $U(1)$ charge conservation.
    (c) The vertical split of the $\delta$ tensors into $\delta^u$ and $\delta^d$ and the horizontal split of the $I$ tensors into $L$ and $R$. 
    (d) The uniform infinite tensor network representation consists of local tensor $O$. The $2\times2$ unite cell in the dashed circle accounts for the stripe order.
    (e) Eigenequation for the fixed-point MPS of the 1D transfer operator.
    (f) Two-point correlation function represented by contracting a sequence of channel operators using canonical MPS.}
    \label{fig:tensor}
\end{figure}

Tensor networks have emerged as a powerful and versatile tool for investigating strongly correlated systems. The first step in applying this method is to express the partition function of a classical lattice model as a tensor network. For the $J_1$-$J_2$ XY model with frustrations, it is crucial to construct a representation that encodes the emergent degrees of freedom into local tensors~\cite{Vanhecke2021,Song2022,Song2023kagome}. A recently developed tensor network framework~\cite{Song2023}, designed to handle frustrated spin models with both discrete and continuous symmetries, provides a natural and effective approach to the current case.

To account for the $Z_2$ degrees of freedom, i.e., the chiralities, associated with each plaquette, the partition function is decomposed into a product of local Boltzmann weights as shown in Fig~\ref{fig:tensor} (a)
\begin{equation}
Z=\mathrm{Tr}e^{-\beta H}=\prod_{i}\int\frac{d\theta_i}{2\pi} \prod_{\square} W_{\square},
\end{equation}
where $W_{\square}(\theta_1,\theta_2,\theta_3,\theta_4)=e^{-\beta H_{\square}}$ is a four-legged tensor with continuous $U(1)$ indices depicted by red dots on the lattice sites. And the local Hamiltonian $H_{\square}(\theta_1,\theta_2,\theta_3,\theta_4)$ on each square is defined as
\begin{equation}
H_{\square}=-\frac{J_1}{2}\sum_{\langle i,j\rangle\in{\square}}\cos(\theta_i-\theta_j)+J_2\sum_{\langle\langle i,j\rangle\rangle\in{\square}}\cos(\theta_i-\theta_j).
\end{equation}

We then use a Fourier transformation to bring the local tensor $W_{\square}$ into a discrete basis, as displayed in Fig~\ref{fig:tensor} (b)
\begin{equation}
I_{n_{1},n_{2},n_{3},n_{4}} =\prod_{i=1}^{4}\int \frac{d\theta _{i}}{2\pi }W_{\square}(\theta _{1},\theta _{2},\theta _{3},\theta _{4}) e^{-in_1\theta_1}e^{-in_2\theta_2}e^{-in_3\theta_3}e^{-in_4\theta_4},
\end{equation}
and the corner sharing constraints among four plaquettes result in a four-leg Kronecker delta tensor connecting the $I$ tensors
\begin{equation}
\delta_{n_1+n_2+n_3+n_4,0}=\int\frac{d\theta'}{2\pi}  e^{in_1\theta'}e^{in_2\theta'}e^{in_3\theta'}e^{in_4\theta'},
\label{eq:conserv_delta}
\end{equation}
representing the conservation law of $U(1)$ charges.

To construct a uniform network, we further decompose the $I$ tensors horizontally by SVD
\begin{equation}
I_{n_{1},n_{2},n_{3},n_{4}}=\sum_{n_{5}}L_{n_{1}n_{2},n_{5}}R_{n_{5},n_{3}n_{4}},
\end{equation}
and split the $\delta $ tensors vertically
\begin{equation}
\delta _{n_{1}+n_{2}+n_{3}+n_{4},0}=\sum_{n_{5}}\delta_{n_{1}+n_{2},n_5}^{u}\delta _{n_{3}+n_{4},-n_5}^{d}
\label{eq:vsp_delta}
\end{equation}
as shown in Fig~\ref{fig:tensor} (c). Finally, the partition function is represented as a translational invariant tensor network in Fig~\ref{fig:tensor} (d)
\begin{equation}
Z=\mathrm{tTr}\prod_s O_{n_1,n_2}^{n_3,n_4}(s),
\end{equation}
where “tTr” denotes the tensor contraction over all auxiliary links. The local tensor $O$ is obtained by grouping the component tensors $R$, $\delta^d$, $L$ and $\delta^u$ inside the dotted circle in Fig~\ref{fig:tensor} (c).

In the thermodynamic limit, the fundamental object for the calculation of the partition function is the row-to-row transfer matrix
\begin{equation}
T(\beta, J_1, J_2)=\mathrm{tTr}\left[\cdots O(p) O(q) O(r) O(s) \cdots\right],
\end{equation}
which is analogous to the matrix product operator (MPO) in a 1D quantum chain.

Due to the stripe structure of the ground state, a $2 \times 2$ unit cell is required, as circled by the dashed lines in Fig.~\ref{fig:tensor} (d). The contraction of the tensor network reduces to finding the leading eigenvalue and eigenvectors
\begin{align}
\begin{split}
T(\beta, J_1, J_2)|\Psi(A, B)\rangle&=\Lambda|\Psi(A', B')\rangle,\\
T(\beta, J_1, J_2)|\Psi(A', B')\rangle&=\Lambda'|\Psi(A, B)\rangle,
\end{split}
\label{eq:eigeneq}
\end{align}
as shown in Fig~\ref{fig:tensor} (e), where $|\Psi(A, B)\rangle$ is the leading eigenvectors represented by matrix product states (MPS) comprising a two-site unit cell of local $A$ and $B$ tensors. This set of fixed-point equations can be accurately solved by the multisite-VUMPS algorithm~\cite{Zauner-Stauber2018,Vanderstraeten2019,Nietner2020}. The precision of the MPS approximation is controlled by the auxiliary bond dimension $D$ of local $A$ and $B$ tensors.

Once the fixed-point MPS is obtained, various physical quantities can be accurately calculated within the tensor-network framework. The entanglement properties of the 1D quantum correspondence can be analyzed through the Schmidt decomposition of $|\Psi(A, B)\rangle$. The entanglement entropy is directly calculated from the singular values as
\begin{equation}
S_E = -\sum_{\alpha=1}^{D}s_{\alpha}^2 \ln s_{\alpha}^2,
\end{equation}
where its singularities serve as a precise criterion for identifying phase transitions. The correlation length is also directly determined from the eigenvalues of the transfer matrix
\begin{equation}
\xi = \ln(\lambda_0 / \lambda_1),
\end{equation}
where $\lambda_0$ and $\lambda_1$ are the largest and second-largest eigenvalues from \eqref{eq:eigeneq}.

Local observables can be evaluated by inserting the corresponding impurity tensors into the original tensor network for the partition function. For example, the expectation value of $e^{iq\theta}$ at site $j$ is
\begin{equation}
\langle e^{iq\theta_j}\rangle  = \frac{1}{Z}\prod_i\int\frac{d\theta_i}{2\pi} e^{-\beta E(\{\theta_i\})} e^{iq\theta_j},
\end{equation}
where $E(\{\theta_i\})$ is the energy for a given spin configuration. The integration \eqref{eq:conserv_delta} at site $j$ is modified to
\begin{equation}
\delta_{n_1+n_2+n_3+n_4+q,0}=\int\frac{d\theta'}{2\pi}  e^{in_1\theta'}e^{in_2\theta'}e^{in_3\theta'}e^{in_4\theta'}e^{iq\theta'}.
\end{equation}
This modification introduces an additional unbalanced delta tensor, $\delta^{+q} = \delta_{m+q,n}$, into the original tensor network, which adjusts the charge conservation laws depicted in Fig.~\ref{fig:tensor} (c),
\begin{equation}
\delta_{n_1+n_2+n_3+n_4+q,0}=\sum_{n_5, n_6}\delta_{n_{1}+n_{2},n_5}^{u}\delta_{n_5+q,n_6}\delta _{n_{3}+n_{4},-n_6}^{d}.
\end{equation}

Using the MPS fixed point, the contraction of the tensor network containing the impurity tensors is reduced to a trace of an infinite sequence of channel operators, which can be further squeezed into the contraction of a smaller network. As shown in Fig~\ref{fig:tensor} (e), the two-point correlation function
\begin{equation}
G_q(r) = \langle \cos(q\theta_i-q\theta_j)\rangle = \langle e^{iq\theta_i}e^{-iq\theta_j}\rangle
\end{equation}
is computed as the trace of a train of $D^2 \times D^2$ matrices, where the unbalanced delta tensors $\delta^{+q}$ and $\delta^{-q}$ are absorbed into the internal indices of the matrices. This process leverages the canonical form of the MPS to ensure efficient computation.

\section{Results}\label{sec:result}
\begin{figure}[tbp]
    \centering
    \includegraphics[width=0.99\linewidth]{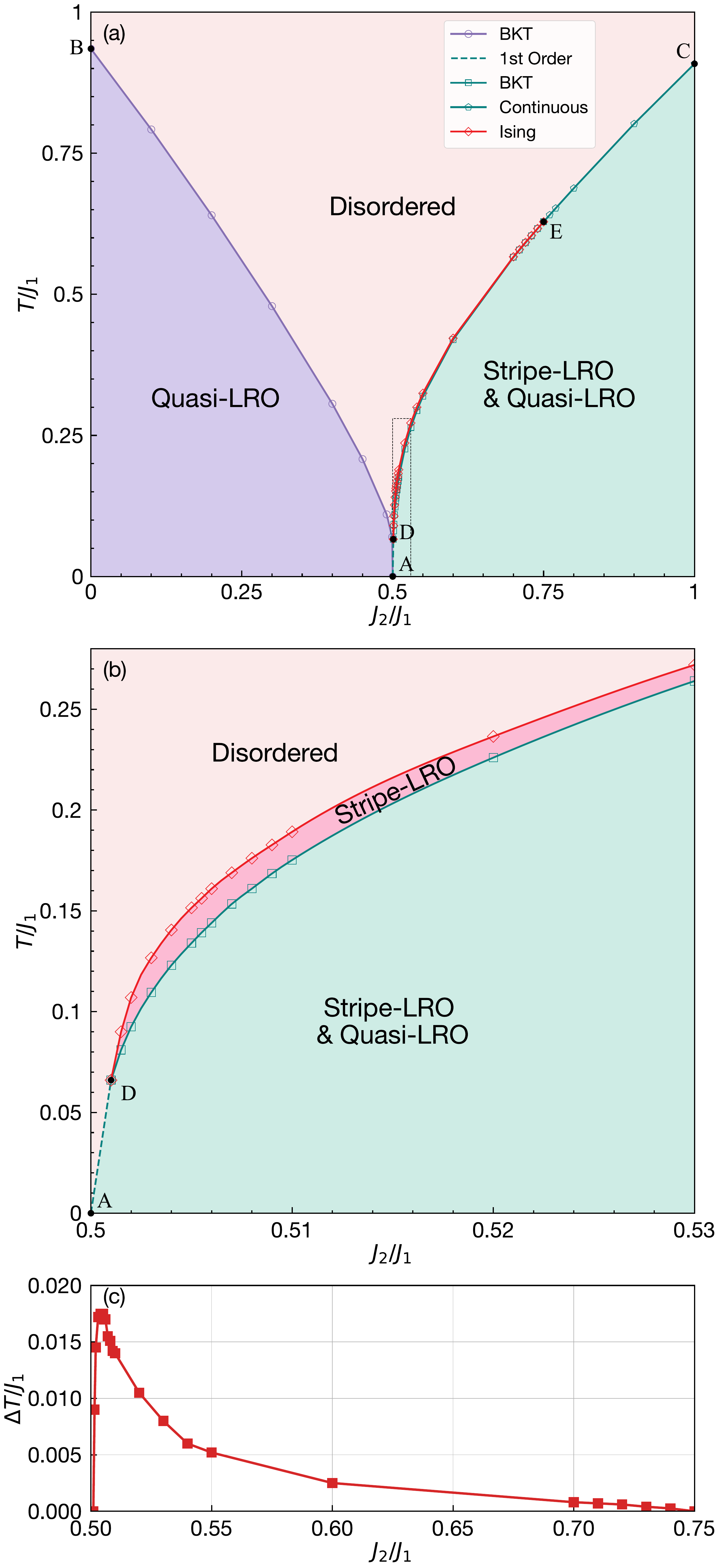}
    \caption{
    (a) The global phase diagram of the $J_1$-$J_2$ XY model. For $J_2/J_1<0.5$, there is a single BKT transition line $AB$. For $J_2/J_1>0.5$, the phase boundary depends on the coupling ratio. As the temperature increases, the ground state $Z_2$ and $U(1)$ order is destroyed through a first-order phase transition $AD$, a crescent-shaped intermediate phase, and a continuous phase transition $EC$, respectively.
    (b) Zoom-in of the phase diagram in the vicinity of $J_2/J_1=0.5$ for a better presentation of the intermediate phase, which starts from point $D$ at $(0.501, 0.066)$ and terminates at point $E$ of $(0.75, 0.628)$. The intermediate phase is characterized by a long-range stripe order in the absence of phase coherence in XY spins, with an upper boundary of Ising transition and a lower boundary of BKT transition.
    (c) The width of the intermediate phase is greatly enlarged compared to the previous results near $J_2/J_1=0.7$.  }
    \label{fig:phase}
\end{figure}

\subsection{Global phase diagram}
The global phase diagram of the $J_1$-$J_2$ XY model has been determined, as shown in Fig.~\ref{fig:phase} (a).  All phase boundaries are determined by the singular behavior of the entanglement entropy $S_E$ of the fixed-point MPS with bond dimension $D=100$ in the thermodynamic limit. The overall structure at high temperatures is consistent with previous studies~\cite{Loison2000}. For $J_2/J_1<0.5$, NN ferromagnetic interactions dominate. There is a BKT transition line $AB$ separating the low-temperature quasi-LRO phase from the hight temperature disordered phase. 

For $J_2/J_1 > 0.5$, the low-temperature phase exhibits long-range stripe order induced by the ``order by disorder" effect, coexisting with a quasi-LRO XY order. The phase transition between the low-temperature stripe-LRO phase and the high-temperature disordered phase has not been well studied previously. From extensive numerical calculations, our results reveal the existence of \textit{an emergent intermediate phase} when the coupling ratio is varied around $J_2/J_1 \sim 0.505$. This intermediate phase features two well-separated phase transitions, as shown in Fig.~\ref{fig:phase} (b). The relative expansion of the intermediate phase is approximately 20 times larger than previously hypothesized, enabling a clearer investigation of the nature of the phase transitions. Specifically, the upper transition is found to belong to the Ising universality class, while the lower transition is of the BKT type. The dependence of the width of the intermediate phase, $\Delta T = T_{\text{Ising}} - T_{\text{BKT}}$, on the coupling ratio is shown in Fig.~\ref{fig:phase}(c). The crescent-shaped intermediate phase begins at the bifurcation point $D$ ($J_2/J_1 = 0.501$) and terminates at the merge point $E$ around $J_2/J_1 = 0.75$.

A segment of first-order phase transition $AD$ is found below $J_2/J_1=0.501$. This result is surprising, as such a transition finds no direct counterpart in the Ising-XY model, suggesting a more intricate mechanism. On the other side, for $J_2/J_1 > 0.75$, there is a single continuous transition line, $EC$, which exhibits \textit{nonuniversal behavior}. Along this line, the critical behavior systematically evolves from second-order transitions to the BKT class, consistent with reports on the Ising-XY model~\cite{Granato1991}.

Since the succession of phases crucially depends on the coupling ratio $J_2/J_1$, our results are discussed in distinct regions to clarify the nature of different phases.

\subsection{Phase transitions at $J_2/J_1=0.7$ and $J_2/J_1=1.0$}
\begin{figure}[tbp]
    \centering
    \includegraphics[width=0.99\linewidth]{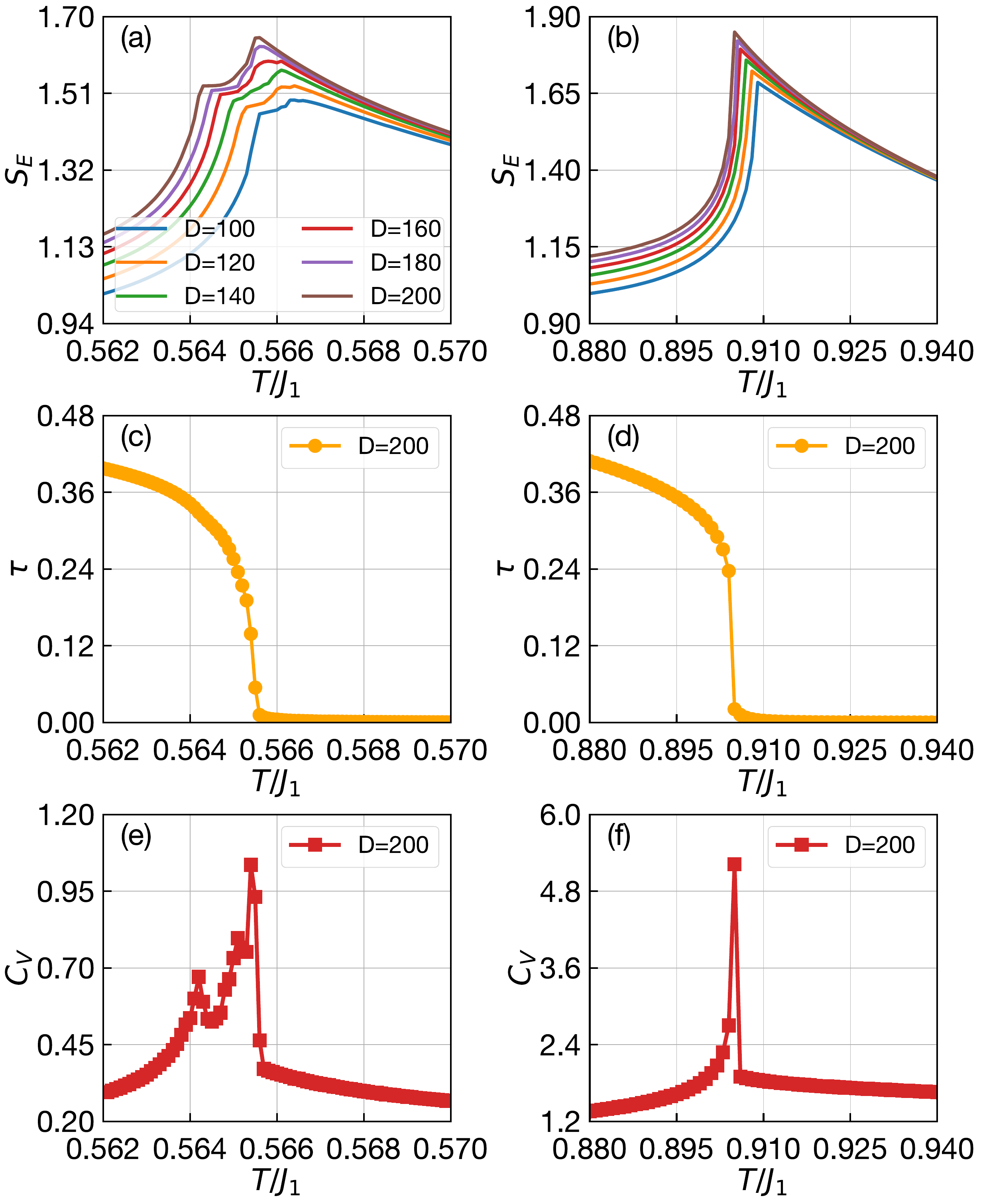}
    \caption{
    (a) The entanglement entropy as a function of temperature exhibits two singularities for $J_2/J_1 = 0.7$.  
    (b) The entanglement entropy displays a single peak for $J_2/J_1 = 1.0$.  
    (c) The stripe order parameter becomes finite below $T_{c2}$ for $J_2/J_1 = 0.7$.  
    (d) The temperature dependence of the stripe order parameter for $J_2/J_1 = 0.7$.  
    (e) The specific heat for $J_2/J_1 = 0.7$ shows a small bump at $T_{c1}$ and a peak at $T_{c2}$.  
    (f) The specific heat for $J_2/J_1 = 1.0$ exhibits a single singularity at $T_c$.}
    \label{fig:J07J1}
\end{figure}

A long-standing debate persists over whether one or two phase transitions separate the low-temperature stripe-LRO phase from the high-temperature disordered phase. Previous Monte Carlo simulations reported conflicting results: a single transition at $J_2/J_1 = 1.0$~\cite{Fernandez1991}, and two closely spaced transitions at $J_2/J_1 = 0.7$~\cite{Loison2000}. In this work, we reconcile these findings by demonstrating that both results are correct and correspond to distinct regions of the phase diagram.

We begin by examining the phase transitions at $J_2/J_1 = 0.7$. As shown in Fig.~\ref{fig:J07J1}(a), the entanglement entropy $S_E$ exhibits two sharp singularities at the critical temperatures $T_{c1}$ and $T_{c2}$, indicating the presence of two distinct phase transitions. Since the positions of these singularities depend on the MPS bond dimension $D$, the critical temperatures $T_{c1}$ and $T_{c2}$ are determined by extrapolating $D$ to infinity. This yields $T_{c1} = 0.5627J_1$ and $T_{c2} = 0.5643J_1$ (see Appendix~\ref{sec:Tc}), in excellent agreement with the critical temperatures reported in~\cite{Loison2000}. The narrow separation between $T_{c1}$ and $T_{c2}$ is characteristic of FFXY models, where the entanglement entropy exhibits a similar structure due to strong proximity effects~\cite{Song2022,Song2023}.

In order to gain insight into the essential physics of different phase transitions, we investigate the thermodynamic properties. To determine the breakdown of $Z_2$, we calculate the stripe order parameter from \eqref{eq:tau}
\begin{equation}
\tau = \frac{1}{4}\langle \cos(\theta_i-\theta_j) +  \cos(\theta_k-\theta_l) - \cos(\theta_i-\theta_l) -  \cos(\theta_j-\theta_k)\rangle,
\end{equation}
within a transition invariant unit cell depicted in Fig.~\ref{fig:tensor} (d). As shown in Fig.~\ref{fig:J07J1} (c), the $Z_2$ order parameter becomes finite at $T_{c2}$, indicating the formation of the long-range stripe order. In addition, we find the expectation value of NN interaction perpendicular to the stripes with alternative signs $\langle \cos(\theta_i-\theta_{i+1})\rangle=(-1)^i|\langle \cos(\theta_i-\theta_{i+1})\rangle|$. 

We then turn to the $U(1)$ symmetry, which is related to the unbinding of vortex-antivortex pairs. A characteristic behavior of the BKT phase transition is a round bump in the specific heat slightly above the transition temperature. The specific heat can be obtained directly from
\begin{equation}
C_V = \frac{du}{dT},
\end{equation} 
where $u$ is the internal energy density obtained from $u=\langle H_{\square}\rangle$. As shown in Fig.~\ref{fig:J07J1} (e), the specific heat exhibits a singularity at $T=T_{c2}$ and a small pointed bump around $T=T_{c1}$. The divergence in specific heat can not fit well with the logarithmic form of the 2D Ising phase transition, and the small bump at the lower temperature is not as smooth as the BKT transition. Such deviation from Ising or BKT universality class is a common issue in FFXY model where two different transitions are close to each other~\cite{Olsson1997,Loison2000}.

\begin{figure}[tbp]
    \centering
    \includegraphics[width=0.99\linewidth]{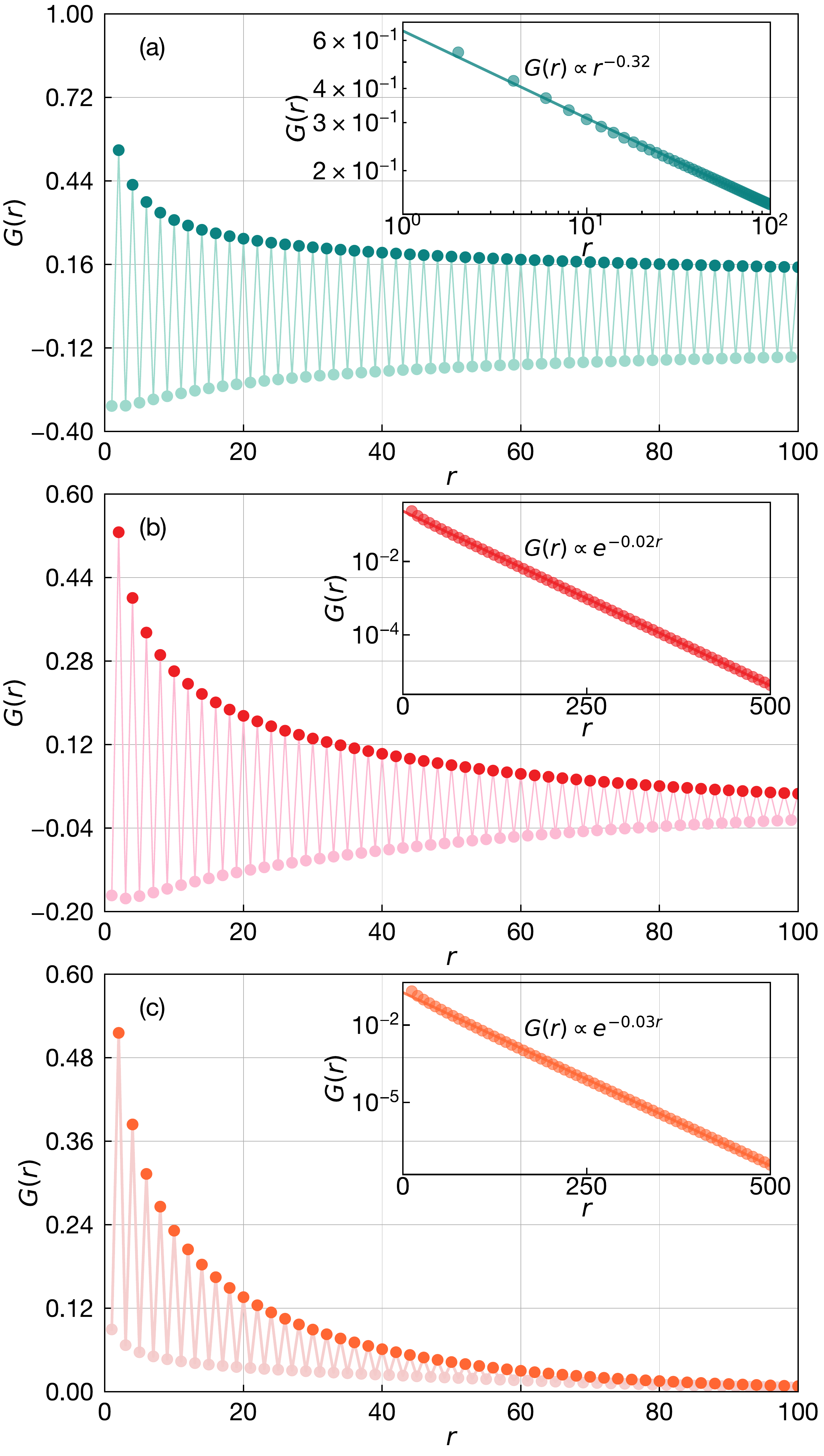}
    \caption{
    Different correlation behaviors along $J_2/J_1 = 0.7$.
    (a) Below $T_{c1}$, at $T/J_1 = 0.562$, the correlation function exhibits power-law decay with oscillations in sign.
    (b) At the intermediate temperature $T/J_1 = 0.565$, between $T_{c1}$ and $T_{c2}$, the correlation function decays exponentially while still oscillating in sign.
    (c) Above $T_{c2}$, at $T/J_1 = 0.568$, the correlation function displays exponential decay with a consistently positive sign.  }
    \label{fig:GrJ07}
\end{figure}

To further explore the nature of the phase transition, we calculate the spin-spin correlation function 
\begin{equation}
G(r)=\langle \cos(\theta_i-\theta_{i+r})\rangle.
\end{equation}
As shown in Fig.~\ref{fig:GrJ07}(a), the spin–spin correlation function $G(r)$ exhibits algebraic decay below $T_{c1}$, indicating the binding of vortices and antivortices. Above $T_{c1}$, however, $G(r)$ decays exponentially, reflecting the loss of phase coherence between vortex pairs. In Fig.~\ref{fig:GrJ07}(b), the exponential decay of $G(r)$ is observed at an intermediate temperature between $T_{c1}$ and $T_{c2}$, further confirming the breakdown of XY order at $T_{c1}$. Moreover, the sign of the correlation function oscillates below $T_{c2}$, signaling the emergence of stripe order. In contrast, as shown in Fig.~\ref{fig:GrJ07} (c), the correlation function remains positive above $T_{c2}$, indicating the destruction of stripe order.

Next, we examine the phase transition at $J_2/J_1 = 1.0$. As shown in Fig.~\ref{fig:J07J1} (b), the entanglement entropy $S_E$ exhibits a single sharp singularity at the critical temperature $T_c$, indicating a single phase transition where both Ising and XY orders are simultaneously lost. By extrapolating to an infinite bond dimension, the critical temperature is determined to be $T_c = 0.9007$, which is in excellent agreement with previous results~\cite{Fernandez1991}.

The $Z_2$ order parameter $\tau$ is shown in Fig.~\ref{fig:J07J1}(d), taking a finite value below $T_c$. As the temperature approaches $T_c$ from below, $\tau$ vanishes continuously following the scaling relation $\tau \sim t^\beta$, where $t = (T_c - T) / T_c$. Notably, the critical exponent $\beta = 0.156$ differs from the $\beta = 1/8$ value characteristic of the 2D Ising model. Furthermore, the specific heat is presented in Fig.~\ref{fig:J07J1}(f), showing a single peak at $T_c$. However, the specific heat curve does not fit a logarithmic scaling and instead displays a discontinuous jump on the high-temperature side. These thermodynamic observations suggest that the system deviates from both the BKT and 2D Ising universality classes due to the close interplay between $U(1)$ and $Z_2$ symmetries, consistent with previous Monte Carlo results~\cite{Fernandez1991}.

As $J_2/J_1$ increases significantly ($J_2/J_1 \gg 1$), the system effectively decouples into two very weakly coupled XY models, and the phase transition is expected to belong to the BKT universality class. Consequently, the universality class along the transition line $EC$ in the phase diagram evolves, corresponding to a specific branch in the Ising-XY model phase diagram. This branch remains controversial, regarding whether the observed variation in critical exponents and central charge represents a genuine physical phenomenon~\cite{Granato1991} or is merely a numerical artifact~\cite{Cristofano2006}. For this reason, we denote the transitions along line $EC$ as a continuous transition line.

\subsection{The intermediate phase}
\begin{figure}[tbp]
    \centering
    \includegraphics[width=0.99\linewidth]{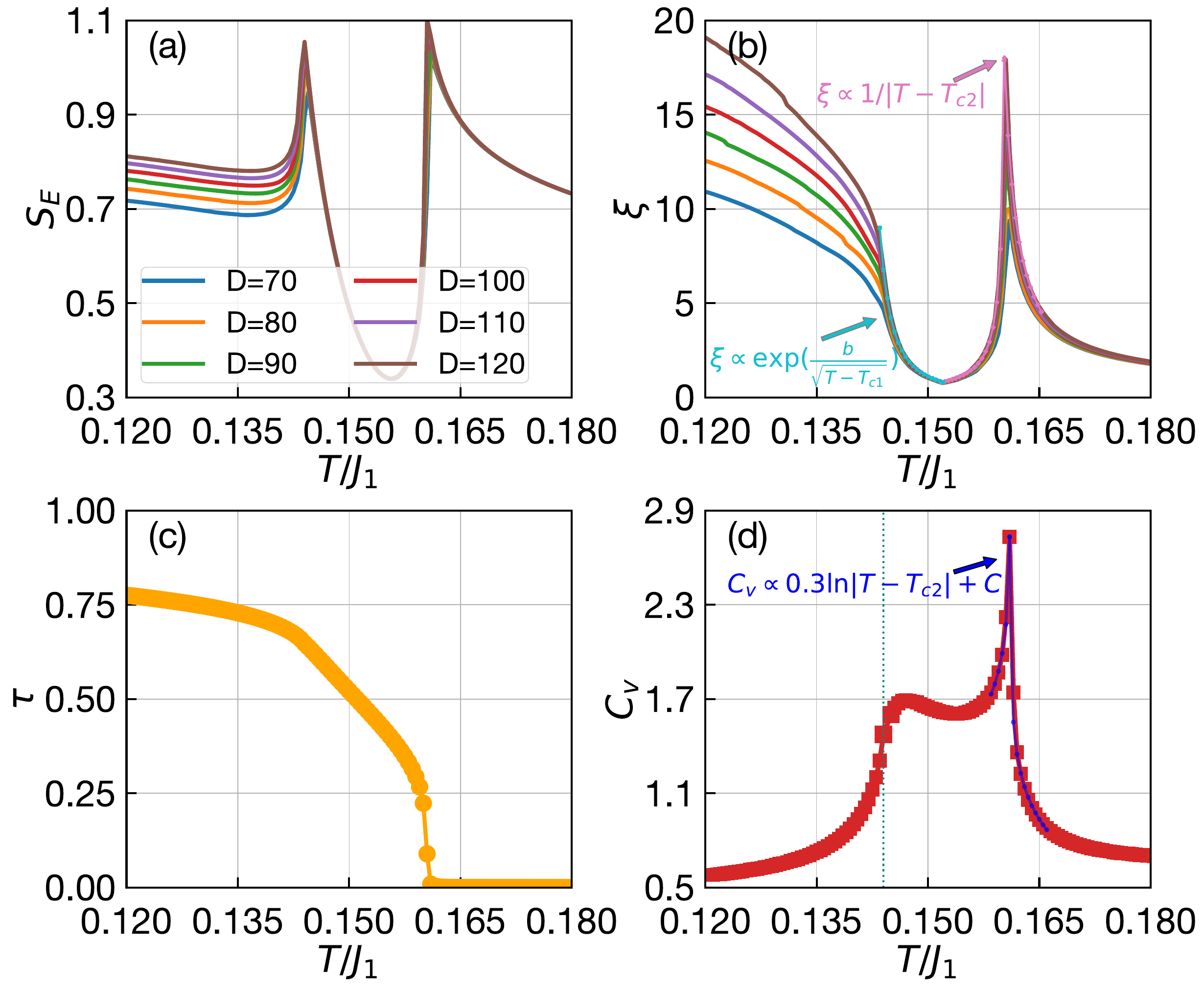}
    \caption{
    (a) The entanglement entropy as a function of temperature, showing well-separated singularities for $J_2/J_1 = 0.506$.  
    (b) The correlation length fitted to BKT and 2D Ising critical behaviors at $T_{c1}$ and $T_{c2}$, respectively.  
    (c) The stripe order parameter exhibiting a two-step feature below $T_{c2}$.  
    (d) The specific heat displaying a bump slightly above $T_{c1}$, characteristic of a BKT transition, and a logarithmic divergence at $T_{c2}$, indicative of a 2D Ising transition. The vertical green dotted line serves as a visual guide for $T_{c1}$.}
    \label{fig:J0506}
\end{figure}

The close proximity of the two phase transitions in frustrated XY models, combined with the lack of sharp thermodynamic signatures, makes distinguishing between them particularly challenging~\cite{Teitel1983,Choi1985,Yosefin1985,Thijssen1990,Santiago1992,Granato1993,LeeJR1994,LeeSy1994,Santiago1994,Olsson1995,Cataudella1996,Olsson1997,Boubcheur1998,Hasenbusch2005,Okumura2011,Nussinov2014,Lima2019,Song2022}. Despite extensive research, the sequence and underlying nature of these transitions remain unclear. Specifically, the specific heat does not align well with the predictions of either the BKT or Ising universality classes, and the critical exponents show some deviations. To gain deeper insight into their nature, it is crucial to reduce the proximity effect by achieving a larger separation between the transitions. We find that, by moving to the region around $J_2/J_1 \sim 0.505$, this separation becomes more pronounced, allowing the Ising and BKT transitions to be clearly identified and verified.

As shown in Fig.~\ref{fig:J0506} (a), along $J_2/J_1 = 0.506$, the entanglement entropy displays two sharp peaks, signaling the presence of two phase transitions at distinct temperatures. The peak positions remain nearly unchanged for large bond dimensions, allowing us to accurately determine the phase boundaries at $T_{c1} = 0.1440J_1$ and $T_{c2} = 0.1605J_1$. The relative separation between the transitions is calculated as $r = (T_{c2} - T_{c1}) / T_{c1} = 11.5\%$, which is significantly larger than the $r < 2\%$ observed in previous studies of the FFXY model. Notably, the largest relative separation for the intermediate phase in our work is approximately $16\%$ at $J_2/J_1 = 0.503$.

The large intermediate phase provides a clear understanding of the nature of the different phase transitions. As shown in Fig.~\ref{fig:J0506} (b), when approaching the critical point $T_{c1}$ from the high-temperature side, the correlation length is well described by an exponentially divergent form:
\begin{equation}
\xi(T) = \exp\left(\frac{b}{\sqrt{T - T_{c1}}}\right), \quad T \to T_{c1}^+,
\end{equation}
where $b$ is a nonuniversal positive constant. This behavior is a hallmark of the BKT transition. Further decreasing the temperature below $T_{c1}$, the correlation length saturates due to the finite bond-dimension effect. In contrast, when approaching $T_{c2}$ from either side, the correlation length diverges as
\begin{equation}
\xi \propto \frac{1}{|T - T_{c2}|},
\end{equation}
which is consistent with the critical behavior of a 2D Ising transition. Additionally, we extract the central charges by fitting the entanglement entropy to the correlation lengths. The central charge is found to be $c = 1$ in the low-temperature phase, indicative of quasi-LRO in XY spins, while it drops to $c = 0$ in the intermediate phase, reflecting the loss of phase coherence above $T_{c1}$ (see Appendix \ref{sec:ccharge}). Moreover, the stripe order parameter exhibits a two-step feature below $T_{c2}$, as shown in Fig.~\ref{fig:J0506}(c). The $Z_2$ order is induced by spin-wave fluctuations and is further enhanced by the quasi-LRO of XY spins that forms below $T_{c1}$.

In most frustrated XY models, distinguishing between different phase transitions using the specific heat is notoriously difficult~\cite{Fernandez1991}. However, in our case, the specific heat $C_V$ exhibits a pronounced bump around $T_{c1}$ and a logarithmic singularity at $T_{c2}$, as shown in Fig.~\ref{fig:J0506} (d). The bump appears slightly above $T_{c1}$, consistent with the behavior typically associated with a BKT transition. Meanwhile, the logarithmic singularity in the specific heat at $T_{c2}$ is a well-known feature of the 2D Ising transition.

From the above analysis, the nature of the intermediate phase is well understood. There are two distinct phase transitions with transition temperatures $T_{c1} < T_{c2}$. The transition at $T_{c2}$ belongs to the 2D Ising universality class, while the transition at $T_{c1}$ is consistent with the BKT universality class. As the system cools, the $Z_2$ symmetry is first broken at $T_{c2}$, marked by the formation of a stripe LRO in the chiralities. At the lower temperature $T_{c1}$, the BKT transition occurs, characterized by the algebraic correlations between vortex-antivortex pairs.

\subsection{The first-order phase transition}
\begin{figure}[tbp]
    \centering
    \includegraphics[width=0.99\linewidth]{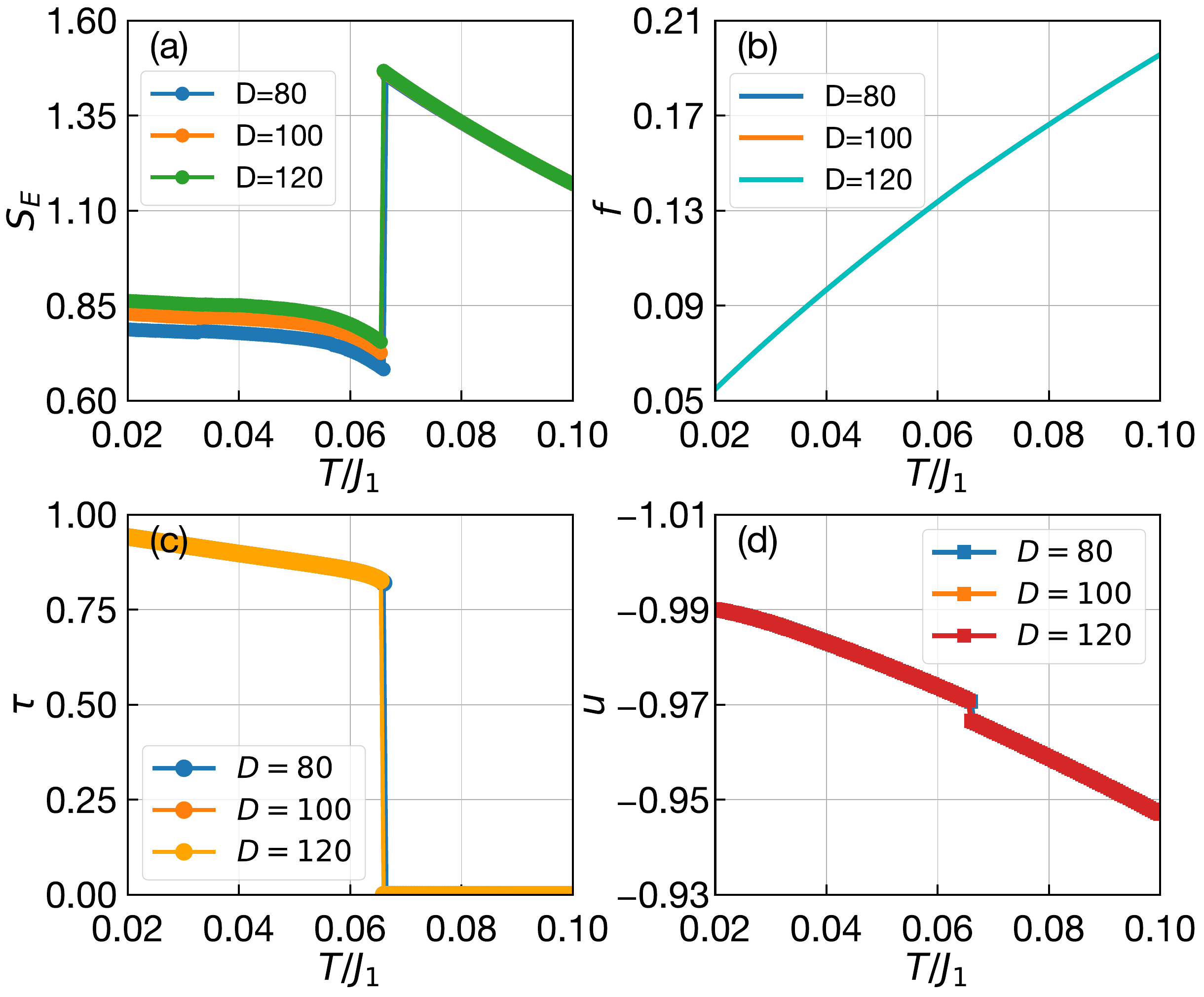}
    \caption{
    Physical quantities for $J_2/J_1=0.501$ at different bond dimensions.
    (a) The entanglement entropy shows a discontinuous jump as a function of temperature, indicating a first-order transition. 
    (b) The free-energy density develops a subtle cusp at the critical temperature. 
    (c) The stripe order parameter exhibits a sharp drop to zero.  
    (d) The internal energy density displays a clear discontinuity.}
    \label{fig:J0501}
\end{figure}

For $J_2/J_1 < 0.501$, a first-order phase transition is identified, based on evidence from both entanglement entropy and thermodynamic quantities. 

As shown in Fig.~\ref{fig:J0501} (a), the entanglement entropy $S_E$ exhibits a discontinuous jump at $T_c = 0.066J_1$ along $J_2/J_1 = 0.501$, signaling a sudden change in the ground state, which is characteristic of a first-order phase transition. In the low-temperature critical phase, $S_E$ increases with the MPS bond dimension $D$. However, thermodynamic quantities remain unchanged for larger $D$, ensuring numerical convergence. As shown in Fig.~\ref{fig:J0501} (b), the free energy density remains continuous across all temperatures but develops a subtle cusp at $T_c$, a hallmark of first-order transitions. Additionally, the stripe order parameter $\tau$, displayed in Fig.~\ref{fig:J0501} (c), undergoes a discontinuous drop to zero at $T_c$, indicating the abrupt loss of stripe order as the system transitions into the disordered phase. Finally, as shown in Fig.~\ref{fig:J0501} (d), the internal energy density exhibits a clear discontinuity at $T_c$, further corroborating the first-order nature of the transition.

The first-order phase transition observed for $J_2/J_1 < 0.501$ in the $J_1$-$J_2$ XY model cannot be fully explained by the mapping to the Ising-XY model described in \cite{Simon1997,Loison2000}. In the Ising-XY model, the branch point marks the merging of separate XY and Ising transitions into a single transition line, which initially corresponds to a segment of continuous transitions (e.g., the transition line $CE$ in the $J_1$-$J_2$ XY model). However, as one moves further along this continuous line, it eventually becomes first-order~\cite{Granato1991,Nightingale1995}. In contrast, in the $J_1$-$J_2$ XY model, as $J_2/J_1 \to \infty$ the system undergoes a BKT transition, which is inherently continuous. This suggests that the first-order transition observed for $J_2/J_1 < 0.501$ arises from a different mechanism from the Ising-XY model.

We should also point out that, due to the finite MPS bond dimension used in our numerical simulations, we cannot rule out the possibility of a continuous phase transition. Further investigation is required to fully understand the origin and nature of this transition.

\section{Conclusion}\label{sec:conclusion}
In this work, we have systematically explored the finite-temperature phase diagram of the $J_1$-$J_2$ XY model on a square lattice using a tensor network approach tailored for frustrated spin systems. By analyzing the singularities of the entanglement entropy and the behavior of thermodynamic quantities, we have resolved several longstanding questions regarding the nature of phase transitions in this model.

The most significant finding of our study is the identification of a large intermediate phase for $J_2/J_1 > 0.5$, which is characterized by long-range $Z_2$ stripe order without phase coherence between XY spins. This phase is bounded by two distinct phase transitions: a higher-temperature Ising transition and a lower-temperature Berezinskii-Kosterlitz-Thouless (BKT) transition. Notably, the relative separation between these transitions becomes significantly larger near $J_2/J_1 \sim 0.505$, providing a unique opportunity to study the interplay between discrete $Z_2$ symmetry breaking and continuous $U(1)$ symmetry. The emergence of this large intermediate phase highlights the crucial role of frustration and thermal fluctuations in stabilizing exotic phases in spin systems.
 
For $0.5 < J_2/J_1 < 0.501$, we identify a narrow region of first-order phase transitions. These are marked by abrupt discontinuities in the stripe order parameter, internal energy density, and entanglement entropy. This first-order transition cannot be fully explained by the conventional mapping to the Ising-XY model, suggesting the need for a more refined theoretical framework. Interestingly, a similar narrow region of first-order transitions has been observed in the $J_1$-$J_2$ Ising model~\cite{Jin2012}, whose underlying nature remains an open question. Exploring a $q$-state $J_1$-$J_2$ model, where $U(1)$ symmetry emerges for $q > 4$~\cite{Lapilli2006}, may provide a systematic understanding of these transitions and their relation to the intermediate phase.

As $J_2/J_1 \to \infty$, the universality of the transition evolves into the BKT class, further emphasizing the rich interplay between discrete and continuous symmetries in this model. An intriguing aspect of this interplay is the conformal anomaly, which reflects the coupling between Ising and XY degrees of freedom. This phenomenon, previously studied in the Ising-XY model~\cite{Nightingale1995,Cristofano2006}, could be further explored and clarified in the context of the $J_1$-$J_2$ XY model through future investigations.

Resolving the finite-temperature phase transitions in fully frustrated XY models remains challenging due to the close proximity of the chiral $Z_2$ symmetry-breaking transition and the BKT transition associated with the $U(1)$ symmetry. In contrast, the $J_1$-$J_2$ XY model, with its larger separation between these transitions for $J_2/J_1 \sim 0.505$, offers a promising framework for studying the intermediate phase identified in this work. These distinctive features make it an appealing candidate for experimental exploration, such as Josephson junction arrays~\cite{Martinoli2000,Cosmic2020}, ultracold atoms~\cite{Kennedy2015,Struck2013}, and optical lattices~\cite{Struck2011,Ozawa2023}, enabling systematic investigations into the interplay between discrete and continuous symmetries.

\begin{acknowledgments}
This work is supported by JSPS KAKENHI Grant No. 23H01092 (2023SY-2027SY) Grant-in-Aid for Scientific Research (B). Part of the computation in this work has been done using the facilities of the Supercomputer Center, the Institute for Solid State Physics, the University of Tokyo.

\end{acknowledgments}

\appendix
\counterwithin{figure}{section}

\section{Order by disorder in the $J_1$-$J_2$ Model}\label{sec:spinwave}
We expand the Hamiltonian around the ground state configuration $\theta_i + \delta \theta_i$, where $\delta \theta_i$ represents small fluctuations.
Expanding $\cos(\theta_i - \theta_j+\delta \theta_i - \delta \theta_j)$ to quadratic order in fluctuations:
\begin{equation}
    \cos(\theta_i - \theta_j) - \sin(\theta_i - \theta_j) (\delta \theta_i - \delta \theta_j) - \frac{1}{2} \cos(\theta_i - \theta_j) (\delta \theta_i - \delta \theta_j)^2.
\end{equation}
The linear term vanishes because the ground state minimizes the Hamiltonian. Thus, the fluctuation contribution is:
\begin{equation}
   - \frac{1}{2} \cos(\theta_i - \theta_j) (\delta \theta_i - \delta \theta_j)^2.
\end{equation}

For NN interactions, the relative angle $\theta_i - \theta_j$ depends on the direction:
\begin{itemize}
    \item Horizontal bonds ($x$-direction): $(\theta_i - \theta_j) = \phi$.
    \item Vertical bonds ($y$-direction): $(\theta_i - \theta_j) = \phi + \pi$, where $\cos(\phi + \pi) = -\cos(\phi)$.
\end{itemize}

Combining the contributions from horizontal and vertical bonds, the onsite terms of $\delta \theta_i^2$ cancel out and the total NN fluctuation contribution is:
\begin{equation}
    \mathcal{A}_\text{NN}^\text{fluct} = - J_1 \sum_{\langle i,j\rangle}\cos(\theta_i - \theta_j) \delta \theta_i \delta \theta_j,
\end{equation}
where $(\theta_i - \theta_j)$ depends on the bond orientation.

For NNN interactions, the relative angle between spins is $(\theta_i - \theta_j) = \pi$. The fluctuation contribution becomes:
\begin{equation}
     \mathcal{A}_\text{NNN}^\text{fluct} = \frac{J_2}{2} \sum_{\langle\langle i,j\rangle\rangle}(\delta \theta_i - \delta \theta_j)^2=2J_2\sum_i\delta\theta_i^2-J_2\sum_{\langle\langle i,j\rangle\rangle}\delta \theta_i \delta \theta_j.
\end{equation}

Therefore we get the total quadratic approximation of the thermal fluctuation
\begin{equation}
     \mathcal{A}^\text{fluct}=2J_2\sum_i\delta\theta_i^2- J_1 \sum_{\langle i,j\rangle}\cos(\theta_i - \theta_j) \delta \theta_i \delta \theta_j-J_2\sum_{\langle\langle i,j\rangle\rangle}\delta \theta_i \delta \theta_j.
\end{equation}

Using the Fourier transformation 
\begin{equation}
\delta\theta_i = \frac{1}{\sqrt{N}} \sum_{\bm{k}} \delta\theta_{\bm{k}} e^{i\bm{k} \cdot \bm{r}_i},
\end{equation}
we get
\begin{align}
\begin{split}
\mathcal{A}^\text{fluct} &= \sum_{\bm{k}} \big[ 2J_2(1-\cos k_x\cos k_y) \\
&\quad\quad\quad - J_1 \cos \phi (\cos k_x - \cos k_y)\big] |\delta\theta_{\bm{k}}|^2 \\
&= N \int \frac{dk^2}{(2\pi)^2} \frac{A_{\bm{k}}}{2} |\delta\theta_{\bm{k}}|^2.
\end{split}
\end{align}

From the Gaussian integral over the spin-wave modes, the entropy contribution can be evaluated as~\cite{Kawamura1984,Henley1989}
\begin{align}
S_0(\phi) &= \text{const.} - \int\frac{dk^2}{(2\pi)^2} \ln A_{\bm{k}}(\phi)\\
&= \text{const.} - \int\frac{dk^2}{(2\pi)^2} \ln\big[(1-\cos k_x\cos k_y)\\
&\hspace{10em}-x(\cos k_x - \cos k_y)\big]\notag,
\end{align}
where we introduce $x=\frac{J_1\cos\phi}{2J_2}<1$. We further expand 
\begin{align}
&\ln\big[(1-\cos k_x\cos k_y)-x(\cos k_x - \cos k_y)\big] \notag \\
=& \ln(1-\cos k_x\cos k_y)\label{eq:g0}\\ 
&- \frac{\cos k_x - \cos k_y}{1-\cos k_x\cos k_y}x \label{eq:g1} \\
&-\frac{(\cos k_x - \cos k_y)^2}{2(1-\cos k_x\cos k_y)^2}x^2 + \mathcal{O}(x^3)\label{eq:g2}.
\end{align}

The linear term \eqref{eq:g1} is an odd function that vanishes upon integration. The integration over \eqref{eq:g1} and \eqref{eq:g2} gives $0.220+0.318x^2$, which means the colinear ordering with $\cos(\phi)=\pm1$ is selected.

\section{Fitting of the critical temperature at $J_2/J_1=0.7$ and $J_2/J_1=1.0$}
\label{sec:Tc}
\begin{figure}[tbp]
    \centering
    \includegraphics[width=0.99\linewidth]{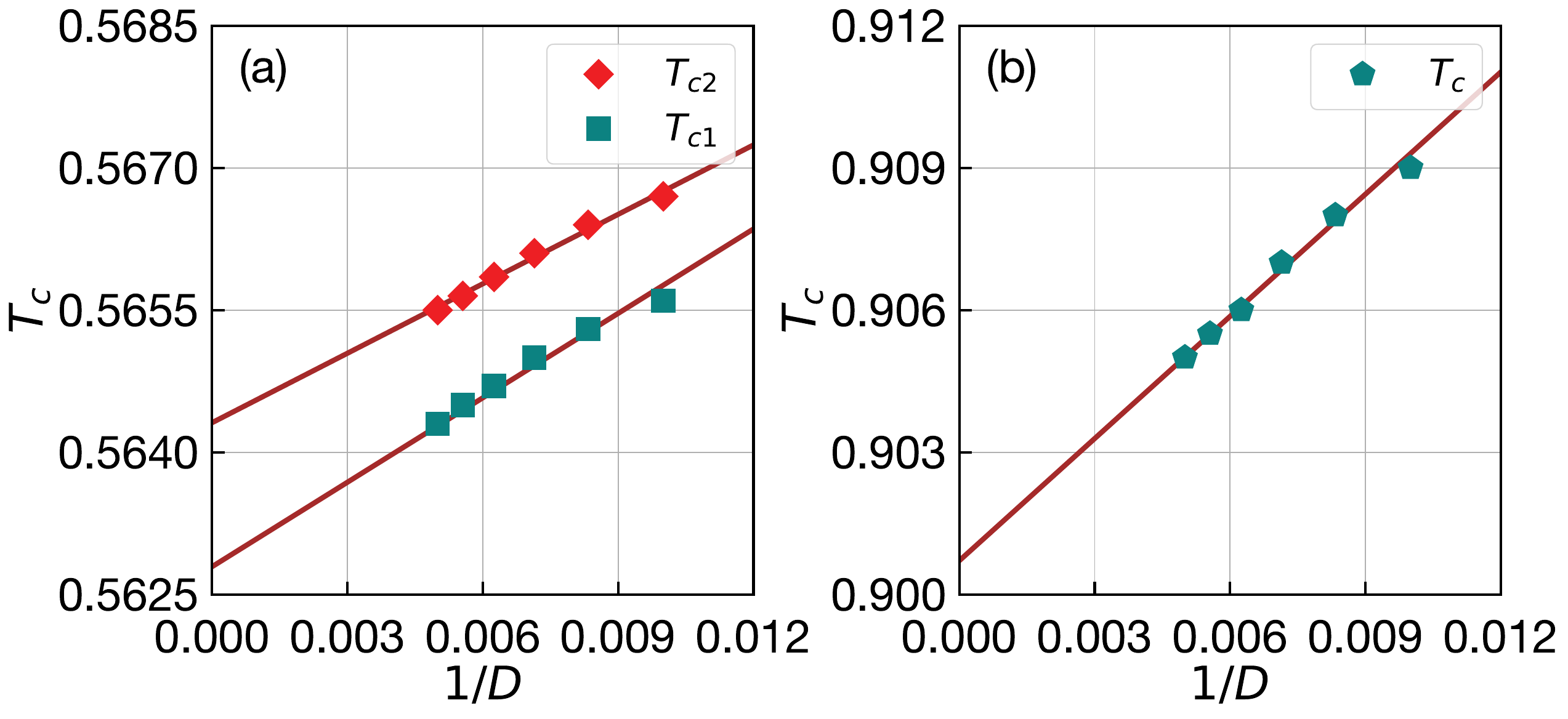}
    \caption{
    The singularity temperatures of the entanglement entropy fitted for MPS bond dimensions from $D=100$ to $200$ with an interval of $20$ with a linear variance of $1/D$. 
    (a) At $J_2/J_1=0.7$, the lower transition temperature is fitted as $T_{\text{c1}}(D=\infty)=0.5627$ and the higher transition temperature $T_{\text{c2}}(D=\infty)=0.5643$.
    (b) At $J_2/J_1=1.0$, the fitting for the transition temperature $T_{\text{c}}(D=\infty)=0.9007$.}
    \label{fig:Tc}
\end{figure}

Due to the finite bond effect, the singularity positions of the entanglement entropy $S_E$ vary with the MPS bond dimension $D$, as shown in Fig.~\ref{fig:J07J1} (a) and (b). Since the finite bond dimensions of the fixed-point MPS can be regarded as a finite cutoff on the diverging correlation length, it is reasonable to use the $1/D$ scaling to extrapolate the critical temperatures. 

The fitting of the critical temperatures for $J_2/J_1=0.7$ is displayed in Fig.~\ref{fig:Tc} (a). The separation between $T_{\text{c1}}$ and $T_{\text{c2}}$ gets larger as the bond dimension increases, which indicates that large bond dimensions are necessary to clarify the nature of the transitions. Our extrapolation gives $T_{\text{c1}}=0.5627$ and $T_{\text{c2}}=0.5643$ for infinite bond dimensions, which is consistent with previous Monte Carlo results~\cite{Loison2000}.

For $J_2/J_1=1.0$, the entanglement entropy SE develops only one sharp singularity, indicating the simultaneous loss of Ising and XY orders at the same transition. The peak positions are slightly changed with different MPS bond dimensions ranging from $D=100$ to $200$. As shown in Fig.~\ref{fig:Tc} (b), the transition temperature $T_{\text{c}}=0.9007$ is determined by extrapolation, which is in good agreement with early Monte Carlo simulations~\cite{Fernandez1991}.

\section{Central charges}\label{sec:ccharge}
\begin{figure}[tbp]
    \centering
    \includegraphics[width=0.99\linewidth]{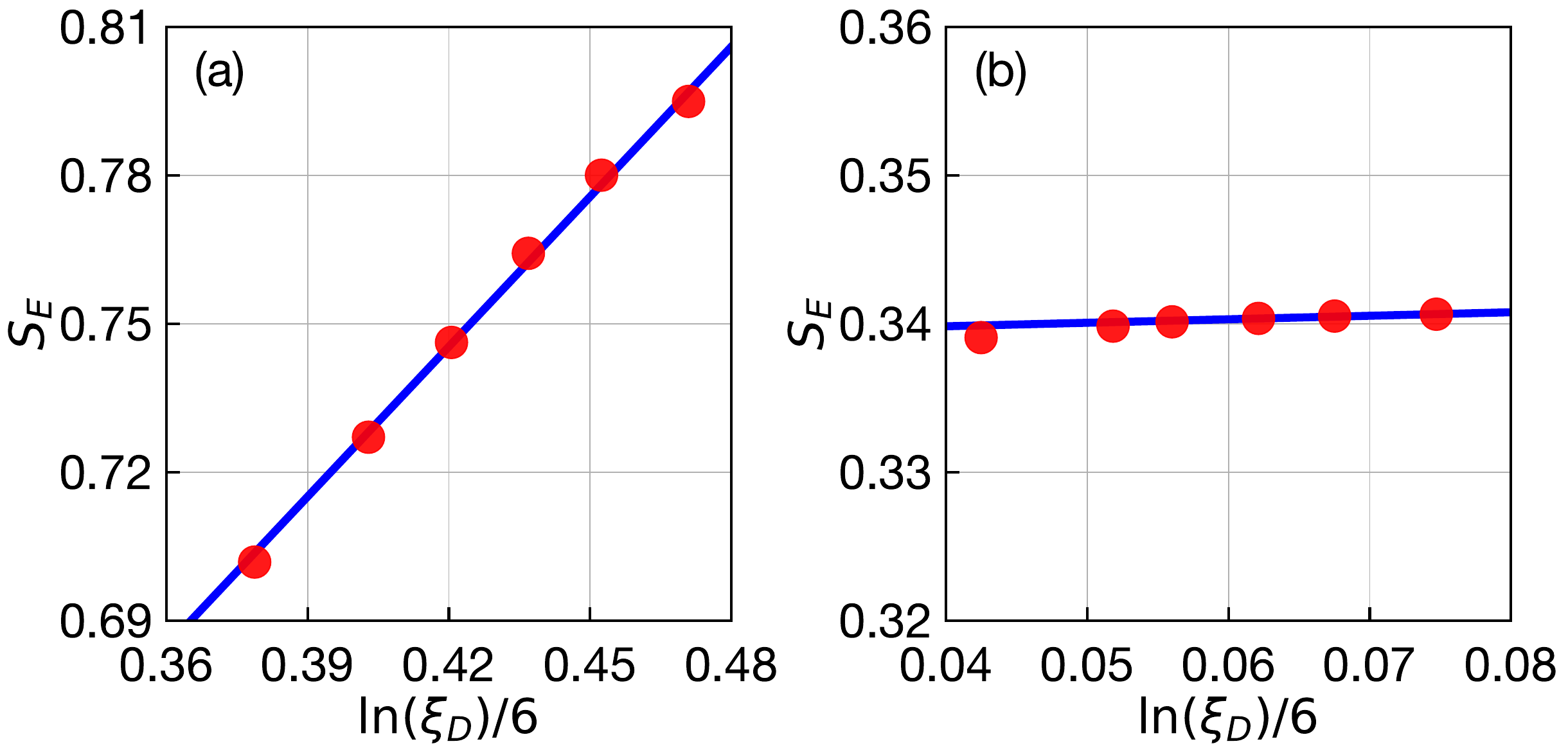}
    \caption{
    Cenral charges extracted from the linear fit of the entanglement entropy with MPS bond dimension from $100$ to $200$ along $J_2/J_1=0.506$. 
    (a) The central charge $c\simeq1.01$ at $T=0.128J_1$ in the low-temperature quasi-LRO phase.
    (b) The central charge $c\simeq 0.02$ at $T=0.156J_1$ in the intermediate stripe-LRO phase.}
    \label{fig:ccharge}
\end{figure}

Within the framework of tensor networks, the central charge of the corresponding conformal field theory can be readily obtained within the critical phases. It is well established that there is a linear scaling relationship between the entanglement entropy $S_E$ and the logarithmic MPS correlation length $\xi_D$ under varying bond dimensions $D$ of the fixed-point MPS~\cite{Calabrese2004}
\begin{equation}
S_E = \frac{c}{6} \ln(\xi_D / a),
\end{equation}
where $a$ is a short length scale.

In Fig.~\ref{fig:ccharge} (a) and (b), we present clear evidence that the central charge is equal to $1$ in the low-temperature quasi-LRO phase, consistent with conformal field theory predictions for superfluid fields. In contrast, the central charge drops to zero in the intermediate phase, reflecting the loss of XY order. It is worth noting that as the system approaches the critical transition point, extracting the central charge from the entanglement entropy and correlation length becomes increasingly challenging. Therefore, we only provide determinations of the central charges in regions sufficiently away from the transition point to ensure reliable results.

\bibliography{ref}

\end{document}